\documentclass[12pt]{article} %latex
\usepackage{latexsym}
\usepackage{amsmath}
\usepackage{pstricks}
\def\ad{^\dagger }
\def\becs{\begin{cases}}
\def\bem{\begin{matrix}}
\def\besp{\begin{split}}

\def\cln{\colon\!} %Symmetrical colon in math expressions

\def\encs{\end{cases}}
\def\enm{\end{matrix}}
\def\ensp{\end{split}} 
\def\hf{{\textstyle\frac{1}{2} }}

\def\hquad{\mspace{8 mu}} %Nominally a 'half quad'; exact half is 9 mu

\def\lgl{\langle }

\def\od{\odot }
\def\ot{\otimes }

\def\rgl{\rangle }
\def\st{\sqrt{2}}

\def\vb{\,|\,}

\def\becs{\begin{cases}}
\def\bem{\begin{matrix}}
\def\besp{\begin{split}}
\def\encs{\end{cases}}
\def\enm{\end{matrix}}
\def\ensp{\end{split}}

\def\scriptl #1 {{\cal #1}}

\def\AS{\scriptl A }
\def\BS{\scriptl B }

\def\ES{\scriptl E }
\def\FS{\scriptl F }

\def\HS{\scriptl H }

\def\al{\alpha }
\def\bt{\beta }

\def\dl{\delta }
\def\Dl{\Delta }

\def\th{\theta }

\def\lm{\lambda }

\def\sg{\sigma }

\begin{document}

\title{The Nature and Location of Quantum Information}

%latex version:
\author{Robert B. Griffiths
\thanks{Electronic mail: rgrif@cmu.edu}\\ 
Department of Physics,
Carnegie-Mellon University,\\
Pittsburgh, PA 15213, USA}

\date{Version of 13 March 2002}

\maketitle  % latex

\begin{abstract}
	Quantum information is defined by applying the concepts of ordinary
(Shannon) information theory to a quantum sample space consisting of a single
framework or consistent family.  A classical analogy for a spin-half particle
and other arguments show that the infinite amount of information needed to
specify a precise vector in its Hilbert space is not a measure of the
information carried by a quantum entity with a $d$-dimensional Hilbert space;
the latter is, instead, bounded by $\log_2 d$ bits (1 bit per qubit).  The two
bits of information transmitted in dense coding are located not in one but in
the correlation between two qubits, consistent with this bound.
	A quantum channel can be thought of as a \emph{structure} or collection
of frameworks, and the physical location of the information in the individual
frameworks can be used to identify the location of the channel. Analysis of a
quantum circuit used as a model of teleportation shows that the location of the
channel depends upon which structure is employed; for ordinary teleportation it
is not (contrary to Deutsch and Hayden) present in the two bits resulting from
the Bell-basis measurement, but in correlations of these with a distant qubit.
In neither teleportation nor dense coding does information travel backwards in
time, nor is it transmitted by nonlocal (superluminal) influences.
	It is (tentatively) proposed that \emph{all} aspects of quantum
information can in principle be understood in terms of the (basically
classical) behavior of information in a particular framework, along with the
framework dependence of this information. 
\end{abstract}

%\maketitle  %revtex

	\section{Introduction}
\label{sct1}

	Quantum information theory has attracted a great deal of attention,
with the interest driven in no small part by potential applications to quantum
cryptography and quantum computing.  It represents an extension of classical
information theory, as developed by Shannon and his successors, to a domain in
which quantum effects are important and therefore quantum mechanics must be
used as the underlying physical theory. It has been suggested, \cite{BeSh98},
that quantum information generalizes classical information in a sense analogous
to that in which complex numbers generalize real numbers.  However, despite
impressive advances in the field, the specifics of this generalization remain
unclear, and various experts have presented rather different views as to the
nature of quantum information; see, for example, \cite{Jzs98}, \cite{DtHy00},
and
\cite{Wer01}.  It would seem that the organizing principles and general laws
which govern quantum information have yet to be discovered, or at least have
yet to be recognized for what they are.

	The aim of this paper is to make a clear, specific proposal about the
nature of quantum information, and then apply it to the problem of the physical
location, in space and time, of information in a quantum system. The proposal
may or may not turn out to be correct (or interesting, or useful, etc.), but it
is at least unambiguous, and can be tested in various ways. In this paper we
apply it to some specific examples, including dense coding and teleportation,
to illustrate what the concepts mean, and to show that they provide sensible
results in certain situations which have often seemed obscure and hard to think
about in physical terms, even though their mathematical description is
relatively straightforward.

	Our presentation begins in Sec.~\ref{sct2} with a brief discussion of
classical information and its location in space and time.  This shows what it
is that we want to generalize to the quantum case, and provides some analogies
which are helpful later in the paper.  The discussion of quantum information
begins in Sec.~\ref{sct3} with an analysis of how much information can be
contained in, or carried by, a single spin-half particle, a single qubit. We
argue that this is at most one bit, and give reasons why it is not larger
despite the fact that an infinite amount of information is needed to describe
(with infinite precision) the mathematical state if the qubit as an element of
its Hilbert space.  In addition, in Sec.~\ref{sct3b} we develop a classical
analogy for a spin-half particle, one that provides what seems to be a more
satisfactory intuitive picture than the fairly common mental image of a
gyroscope with its axis pointed in a precise direction in space.

	Since the quantitative measures of information, such as entropy, used
in ordinary (classical) information theory are expressed in terms of
probabilities, their extension to the quantum domain requires a consistent
probabilistic formulation of quantum theory.  For our purposes the traditional
approach based upon outcomes of measurements is not adequate, and
Sec.~\ref{sct4a} indicates how this goal can be reached using quantum
frameworks (consistent families). One consequence, Sec.~\ref{sct4b}, is a
maximum information capacity of ``one bit per qubit'' for carriers of quantum
information, confirming the conclusions reached in Sec.~\ref{sct3}.  The use of
frameworks also allows one to assign a location to quantum information, as
explained in Sec.~\ref{sct4c}.

	Dense coding is taken up in Sec.~\ref{sct5}.  It has often been
supposed that this is a situation in which a single qubit can transport two
bits of information.  However, using the tools developed in Sec.~\ref{sct4} we
show that this is not so: the two bits (and each bit independently) reside on
two qubits, not one, so the maximum capacity result of Sec.~\ref{sct4b} is
respected.  We also show that the information does not travel backwards in
time, and argue that it is not transmitted instantaneously, by some peculiar
superluminal influence violating classical causality, to a distant qubit.  To
be sure, there is something mysterious (i.e., nonclassical) about dense coding,
and our analysis makes clear precisely what it is.

	Quantum theory allows a physical system to be described by a variety of
different frameworks, and the presence or absence of information (and hence its
location) depends, in general, upon the framework.  As an aid to thinking about
this
\emph{framework dependence}, which is a novel feature of quantum mechanics with
no classical analog, we introduce in Sec.~\ref{sct6a} the notion of a
\emph{structure}, or collection of frameworks with a common algorithm for
calculating probabilities.  A particular example of a structure is an ideal
quantum channel, discussed in Sec.~\ref{sct6b} from the point of view of the
physical location of information in the different frameworks that constitute
the channel, and this is used to define the location of the channel itself. 

	The tools developed in Sec.~\ref{sct6b} for locating a quantum channel
are applied in Sec.~\ref{sct7} to a particular three-qubit quantum circuit
which has been used to discuss teleportation \cite{BrBC98}. This circuit is
analyzed from different points of view, i.e., by using different structures,
and the location of the channel is shown to depend upon which of these one
employs.  Teleportation itself corresponds to a particular structure, and we
identify the location of the relevant information during the ``classical''
communication process.  Once again, the information does not travel backwards
in time, nor are there magical long-range influences.

	The concluding Sec.~\ref{sct8} summarizes the basic proposal of this
paper in terms of two theses. The first is that the use of frameworks is an
appropriate (consistent, useful, etc.) generalization of classical information
to quantum mechanics: it embodies at least something of what quantum
information is all about.  The second, more speculative thesis is that this
generalization is sufficient to cover, at least in principle, \emph{all} of
quantum information: that is, there is no additional, irreducibly-quantum
information which is not ``classical'' information in a particular
framework, or else the framework dependence of such information.

	The only other work we are aware of which addresses in a similar way
questions about the location of quantum information is that of Deutsch and
Hayden \cite{DtHy00}.  Aside from one or two points of similarity, e.g., we
agree that there can be no instantaneous (superluminal) transmission of
information in a quantum system, their approach is completely different from
ours; see the comments in App.~\ref{apa}.

	\section{Classical Information}
\label{sct2}

	The morning newspaper contains an account of a speech which the
President made yesterday.  One can say that the newspaper \emph{contains
information} because from the shapes and positions of the symbols on the paper
it is possible to infer something of what the President said.  It is a
\emph{carrier} of information about a \emph{message}, in this case the original
speech.  This information has a physical location in the same sense that the
newspaper has a physical location.  Of course this location is not unique ---
there are many copies of the newspaper --- nor is it a mathematically precise
location, since the printed words take up some space, and the message cannot be
recovered from a single letter or even a single word; in some sense it resides
in the correlations among the symbols. 

	The notions of information and its location can be made more precise by
embedding them in a probabilistic or stochastic model, something which is
necessary if one wants to use the concepts of classical information theory
\cite{CoTh91}.  Let $M$ be a random variable representing the different
possible messages which can be produced by a source, and let $N$ be another
random variable representing different physical states of a carrier of
information.  If for any $M$ that occurs with nonzero probability there is some
$N$ such that the conditional probability $\Pr(M\vb N)$ is 1, we shall say that
the carrier contains the information produced by the source, whereas if $M$ and
$N$ are statistically independent, the carrier does not contain this
information.  It is convenient to use Shannon's mutual information $I(M\cln N)$
\cite{CoTh91} as a quantitative measure of the information present in $N$,
especially when, as will be the case in most of the discussion that follows, we
are interested in a small number of messages which occur with equal
probability.

	As a simple example, suppose Alice sends a one bit message $M=0$ or 1
to Bob by mailing him a colored slip of paper, red ($R$) for 0, green ($G$) for
1, inside an envelope. While it is in transit the message is contained in or
carried by the slip of paper, and if 0 and 1 are equally likely, the slip
carries \hbox{$\log_2 2=1$} bit of information.  Next consider a more
complicated arrangement in which the message is encoded in two slips of paper
which Alice sends to Bob in separate envelopes: identical colors, $RR$ or $GG$,
chosen at random, stand for 0, and opposite colors, $RG$ or $GR$, also chosen
at random, stand for 1.  In this case a single envelope contains no information
about the message, since both for $M=0$ and $M=1$, $R$ or $G$ will be present
with probability 1/2.  Instead, the message is present as a \emph{correlation}
between the contents of the two envelopes.  It is \emph{delocalized} in the
sense that it is not present in either envelope by itself, but can still be
said to be in the region of space occupied by both envelopes --- the union or
their interiors, to use a set-theoretical term --- even in cases in which the
envelopes are far away from each other.  Information is something abstract
whose location need not satisfy the usual rules for the location of material
objects.

	A third example useful in a later discussion of an analogous quantum
situation is the following: Suppose that Alice and Bob initially share slips of
paper that are known to be identical in color: both $R$ or both $G$.  For
example, they were prepared and placed in separate envelopes sent to Alice and
Bob by Charlie, who flipped an honest coin to decide between $RR$ and $GG$.
Alice now sends a one bit message $M$ to Bob in the following way. For $M=0$
she sends him the slip of paper in her possession, and for $M=1$ she sends him
a slip of the opposite color.  Where is the information about $M$ located while
this slip is in transit from Alice to Bob?  It cannot be in the slip which is
in the mail, because in our stochastic model both $M=0$ and $M=1$ lead to $R$
and $G$ with equal probability.  Nor is it, obviously, in the slip that is
already in Bob's possession. Instead, it is present in the correlation between
these two slips, and physically located in the union of the two regions that
they occupy.

	One can imagine more complicated situations.  For example, modify the
previous example so that Charlie prepares $RG$ and $GR$ as well as $RR$ and
$GG$, each with probability 1/4, and sends the first slip to Alice and the
second to Bob.  Later, at the same time Alice sends her message to Bob
by following the procedure in the preceding paragraph, Charlie mails to Bob a
yellow slip of paper indicating that the earlier slips he sent were of the same
color, or a blue slip indicating that the colors were opposite.  While the
slips from Alice and Charlie are in transit to Bob, the message is contained in
the correlation between these and the slip already in Bob's possession, whereas
it is not present in the individual slips or in any of the three pairs of two
slips.  A quantum analog of this situation will be discussed later.

	\section{Spin Half}
\label{sct3}

	\subsection{Distinct states}
\label{sct3a}

	The simplest quantum carrier of information is a two-level system or
qubit, which we shall for convenience think of as the spin degree of freedom of
a spin-half particle, although the polarization of a photon would do just as
well.  Suppose a polarization apparatus (polarizer) produces a particle with
its $z$ component of spin angular momentum $S_z$ equal to $+1/2$ or $-1/2$ (in
units of $\hbar$), depending upon the setting 0 or 1 of a switch $M$. This
particle then travels through a field-free region to a Stern-Gerlach measuring
apparatus where the value of $S_z$ is measured, with outcome $N$ equal to 0 or
1 indicated by a suitable pointer.  As there is a perfect correlation between
the polarization setting $M$ and the measurement outcome $N$, the mutual
information $I(M\cln N)$ is $\log_2 2$ or 1 bit, assuming equal a priori
probabilities for $M$.

	It makes sense to suppose that this 1 bit of information is carried by
the spin-half particle during the time between preparation and measurement,
since, for example, the perfect correlation between $M$ and $N$ requires that
the particle actually arrive at the measuring apparatus, and can be altered if
the particle is scattered or passes through a magnetic field during its
trajectory.  That the particle contains information about $M$ can be confirmed
using the formulation in Sec.~\ref{sct2} by working out the joint probability
distribution for the random variables $M$ and $S_z$:
\begin{equation}
  \Pr(M=0,S_z = +1/2) = 1/2 = \Pr(M=1,S_z=-1/2),
\label{eq1}
\end{equation} 
%with \hbox{$\Pr(M=0,S_z = -1/2)$} and $\Pr(M=1,S_z=+1/2)$ both equal to
with $\Pr(M=0,S_z = -1/2)$ and $\Pr(M=1,S_z=+1/2)$ both equal to 0. Introducing
probabilities that refer to a microscopic quantum system requires some care,
and we shall return in Sec.~\ref{sct4a} to the considerations which justify
\eqref{eq1}.  Assuming its validity, the situation is analogous to one in which
a switch setting of $M$ equal to 0 or 1 on one piece of apparatus gives rise to
two distinct macroscopic electrical signals in a cable (the analogs of $S_z=\pm
1/2$) connecting this apparatus to a second one in which the signals give rise
to two different pointer positions.  In such a situation one would not hesitate
to say that the information about $M$ is carried by the electrical signal, and
in the same way it seems plausible that information is carried by the spin half
particle.

	However, the quantum case is conceptually more complicated than its
classical analog because the polarizer can prepare other components of spin
angular momentum besides $S_z$.  There might, for example, be additional switch
settings $M=\bar 0$ or $\bar 1$ which result in a particle with $S_x = +1/2$ or
$-1/2$, respectively.  The vectors in the Hilbert space corresponding to
$S_x=\pm 1/2$ are quite distinct from those corresponding to $S_z=\pm 1/2$, and
if these were to correspond to physically distinct states of the quantum
particle one might plausibly conclude that the particle could carry $\log_2 4 =
2$ bits of information.  Indeed, there is no reason not to allow the polarizer
the possibility of producing a particle with a polarization $S_w=1/2$ for $w$
any direction in space.  Since the states (or, to be more precise, the rays or
one-dimensional subspaces) of the Hilbert space corresponding to distinct $w$
are different, a single spin-half particle would, according to this point of
view, be capable of containing or carrying an infinite amount of information.
(This idea seems to be widespread; for a specific reference, 
see \cite{Jzs98}.)

	There are, however, serious objections to the possibility of a spin-half
particle carrying more than 1 bit of information in its spin degree of freedom.
To begin with, the classical definition of $\log_2 n$ as the information
(entropy) of a source of $n$ equally-likely messages requires that these
messages be \emph{distinct} from one another.  From a quantum perspective, any
two states of affairs which are classically (i.e., macroscopically) distinct,
such as two pointer positions, always correspond to quantum states that are
\emph{orthogonal} to each other.  (If the distinct states are described by two
density matrices $\rho$ and $\rho'$, then $\rho\rho'=0$.)  For spin half, the
eigenstates corresponding to $S_z=+1/2$ and $-1/2$ are orthogonal, whereas
eigenstates of $S_z$ are not orthogonal to eigenstates of $S_x$. Therefore
$S_z=+1/2$ and $S_x=+1/2$ do not correspond to distinct states in the classical
sense, and this is confirmed by the fact that there is no measurement that can
distinguish them.  To be sure, there might be a distinction which is physically
present but is inaccessible to measurements.  But this sounds a bit like a
student's claim to understand a subject when he has just failed the
examination, and one tends to be skeptical.

	To be sure, the states $S_z=1/2$ from $S_x=1/2$, while they cannot be
distinguished for an individual particle, are distinct in that they give rise
to different statistical predictions for the outcomes of various measurements.
Such a distinction does not, however, mean that individual particles are in
some sense distinct depending on whether they were prepared with the polarizer
switch setting of $M=0$ or $M=\bar 0$, as can be seen from a classical analogy.
Consider an ensemble of bolts produced by machine No.~1 that creates one
defective bolt in 10, and another ensemble produced by machine No.~2 that
creates one defective bolt in 20.  Separate ensembles of bolts produced by the
two machines can be distinguished with a probability that approaches 1 if the
ensembles are sufficiently large, but the distinction cannot be made on the
basis of a single bolt.  Distinct probability distributions do not mean
distinct bolts, and it would be somewhat strange to claim that because the
probability of a defective bolt is described by a real number $p$ having an
uncountably infinite number of possibilities, that therefore every bolt carries
an infinite amount of information relative to its being or not being defective.
Of course, a pure quantum states is not the same thing as a probability
distribution.  But when it functions as a pre-probability \cite{nt0} used to
calculate a probability distribution it has a somewhat similar character.

	In summary, the undeniable \emph{mathematical} difference between the
states $S_z=1/2$ and $S_x=1/2$ as elements (or, better, rays) in the Hilbert
space need not reflect a \emph{physical} distinction in any sense analogous to
those between two non-identical classical messages.  And if it does not, then
it provides no basis for supposing that a single qubit can contain more than
one bit of information.

	\subsection{Classical analogy}
\label{sct3b}

	One source of the (mistaken, in our opinion) idea that a spin half
particle can carry an infinite amount of information is a mental image in which
one thinks of it as a classical object like a gyroscope, spinning about an axis
pointing in a well-defined direction in space, say the $+z$ direction if the
quantum state is $S_z=+1/2$.  Such mental images, or models, of quantum objects
in classical terms are probably unavoidable in that quantum theory is a product
of human thinking that is shaped in large part by ordinary ``classical''
experience, and they need cause no harm if one understands their limitations
and realizes how they can be misleading.  The gyroscope model under discussion
is misleading in that the components of angular momentum in directions
perpendicular to the special axis are zero, whereas one knows that the quantum
operators $S_x^2$ and $S_y^2$ are identical to $S_z^2$, and thus thinking of
$S_x$ and $S_y$ as equal to zero when $S_z=1/2$ cannot be correct.  A
modification of this model, which is still classical but a bit less
misleading, has the axis of the gyroscope pointing in a random direction in
space consistent with a definite positive value of $S_z$, but with random
non-zero values of $S_x$ and $S_y$.  It resembles what one finds in older texts
on atomic physics, where even when $S_z$ has its maximum value, the total
angular momentum is shown as a vector pointing in a direction which does not
coincide with the $z$ axis.

	If one uses the first model, in which $S_z=+1/2$ corresponds to the
gyroscope axis pointing along the $z$ axis, one tends to get the idea that the
quantum state contains a large amount of \emph{orientational} information
needed to specify the precise direction in which the spin is ``pointing''.
This is reinforced by the observation that a spin polarizer, a large classical
apparatus, can be oriented quite precisely, combined with the plausible, but
questionable, supposition that the same precise orientation is imparted to the
spin-half particle. The second model, in which the orientation of the gyroscope
axis is to a large extent random, suggests that, on the contrary, while the
polarizer can fix one component of the angular momentum quite precisely, it
does not impart a precise orientation, for the other components of angular
momentum are random.  That the polarizer should produce random values for the
components of angular momentum perpendicular to the polarization direction is
consistent with the usual belief that a measurement of one component of the
spin of a spin half particle randomizes the orthogonal components, when one
remembers that such measurements are one way of producing polarized
particles. In summary, while this second model remains classical, and is thus
bound to be misleading in some respects, in other respects it provides a
significantly better intuitive feeling for the quantum situation than does the
first model.

	It is possible to carry out a quantitative analysis of the information
contained in a simplified version of the second model that no longer respects
the requirement of rotational invariance.  We suppose that a classical object
has $x$, $y$, and $z$ components of angular momentum equal to $+1/2$ or $-1/2$
in some set of units, and thus a set of eight possible states, which can be
denoted by $(\pm,\pm,\pm)$; e.g., $(+,+,-)$ stands for $S_x=+1/2$, $S_y=+1/2$,
$S_z=-1/2$.  We assume that the polarizing apparatus in preparing a state
$S_z=+1/2$ randomly perturbs the $x$ and $y$ components of angular momentum, so
that it produces one of the four states $(\pm,\pm,+)$ with equal probability.

	Now suppose that the polarizer is equipped with six switch settings
allowing it to produce the states $S_x=\pm 1/2$, $S_y=\pm 1/2$, and $S_z=\pm
1/2$, where in each case the state has the specified value for that component
of angular momentum, but random values for the other two.  It is then clear
that from the resulting ``spin configuration'', of the object, say $(+,-,+)$,
one can decide unambiguously between the polarization settings $S_x=+1/2$ and
$S_x=-1/2$, but not between $S_x=+1/2$ and $S_z=+1/2$.  For this model one can
calculate the mutual information between the 6 polarization settings, assumed
equiprobable, and the spin state, and the result is 1 bit.  That is to say, the
object can carry enough information to distinguish two opposite values of one
component of angular momentum, but contains no additional information about the
orientation of the polarizer.  

	Since this model is classical it cannot clarify all the
conceptual difficulties of the quantum case.  However, it does suggest how a
preparation procedure can impart to a single particle much less information
than is required to describe the procedure (in particular, the orientation of
the polarizer) itself.

	\section{Quantum Information}
\label{sct4}

	\subsection{Frameworks}
\label{sct4a}

	While the qualitative arguments in Sec.~\ref{sct3a} and the (second)
classical model in Sec.~\ref{sct3b} provide reasons for supposing that a
spin-half particle cannot carry a large amount of information in its spin
degree of freedom, they are obviously not substitutes for a precise discussion
properly grounded in the fundamental principles of quantum theory, which we
shall now begin to construct.

	Classical information theory is formulated in probabilistic terms, and
hence it makes sense to try and construct a quantum counterpart using a
consistent probabilistic formulation of quantum theory.  Unfortunately, quantum
textbooks usually introduce probabilities in terms of the outcomes of
measurements, and take a more or less agnostic attitude towards what goes on
between measurements, or between the preparation of a system by some
macroscopic device and its measurement by another.  During this time interval
the quantum system is thought of as a black box, or ``smoky dragon''.  Such an
approach is obviously not adequate for addressing questions about the amount of
information carried by a single qubit when it is not being measured.  One might
still make some progress if one could assume that the outcome of a measurement
reveals a property the measured system had before the measurement took place,
but it is precisely this that the traditional approach denies.  Fortunately,
there are other tools available for assigning probabilities in a consistent way
to quantum systems without reference to measurements
\cite{Omn92,GMH93,Omn94,Grf96,Grf98,Omn99,Grf02}.

	Ordinary probability theory is based upon the notion of a \emph{sample
space} of mutually exclusive possibilities, one and only one of which occurs in
any given experimental trial: e.g., the two sides of a coin, or the six sides
of a die.  The quantum counterpart of a sample space is a \emph{framework},
also known as a consistent family or decoherent set.  The simplest example is
an orthonormal basis $\{|a_j\rgl\}$ of states of the quantum Hilbert space, or,
equivalently, the associated collection of projectors $P_j =
|a_j\rgl\lgl a_j|$ which form a decomposition of the identity: 
\begin{equation}
  I=\sum_j P_j,\quad P_j = P_j\ad,\quad P_j P_k = \dl_{jk} P_j.
\label{eq2}
\end{equation} 
It is important that the states be orthogonal, for only then do they correspond
to physically distinct states that are mutually exclusive (see the discussion
in Sec.~\ref{sct3a}).  A framework or quantum sample space for a system at a
single time is always associated with a decomposition of the identity $\{P_j\}$
with the projectors satisfying \eqref{eq2}, but some of these projectors may
project onto subspaces of dimension greater than one, in which case the
decomposition is not associated with a unique orthonormal basis.

	For a spin-half system the orthonormal basis consisting of the
eigenstates of $S_z$ constitutes a framework, and that corresponding to the
eigenstates of $S_x$ a different framework.  These two frameworks are
\emph{incompatible}: it is impossible to combine them into a single sample
space, because one cannot simultaneously assign values to $S_x$ and $S_z$ for a
spin-half particle (there is no ray in the Hilbert space corresponding to the
combined values) \cite{nt1}.  More generally, two frameworks are incompatible
when the projectors corresponding to one decomposition of the identity fail to
commute with the projectors of a different decomposition.  While it is possible
to assign probabilities separately in different frameworks, it is impossible to
combine probabilities associated with incompatible frameworks, because there is
no common sample space. This is an instance of the \emph{single framework rule}
\cite{nt2}, which states that it is meaningless to try and combine two (or
more) quantum descriptions of a system based upon incompatible frameworks.
This rule is important, because when one is analyzing quantum systems it is
easy to be misled into combining results in a manner which is quite acceptable
in classical physics, where incompatible frameworks never arise, but which
leads to quantum paradoxes.

	When one is considering probabilities of events that take place at
different times, as in \eqref{eq1}, it is necessary to employ a framework or
quantum sample space of \emph{histories}, sequences of events at different
times.  Rules for incompatibility of frameworks and for assigning probabilities
to histories in a single framework have been worked out by Gell-Mann and Hartle
\cite{GMH93}, Omn\`es \cite{Omn92,Omn94,Omn99}, and the author \cite{Grf96};
for a recent, detailed formulation see Ch.~10 of \cite{Grf02}.  These are more
complicated than those for a single Hilbert space at a single time, but for the
purposes of the present paper these complications, embodied in what are called
``consistency conditions'' or ``decoherence conditions'', can for the most part
be ignored; exceptions will be noted as they occur.
	One can also employ frameworks to discuss measurements, provided the
measuring apparatus is regarded as a quantum object, part of an overall system
that is described using quantum theory. The probabilities of measurement
outcomes are then the same as those calculated by the rules of standard
textbook quantum theory.  Hence the use of frameworks is not a new version of
quantum theory different from what is found in textbooks, but instead a
consistent extension that allows probabilistic descriptions of microscopic as
well as macroscopic systems.

	The use of frameworks provides a quantum mechanical justification for
\eqref{eq1}, where the values of $M$ refer to the polarizer switch setting at a
time before the particle is produced, and $S_z=\pm1/2$ to a later time before
any measurement occurs.  The assignment of probabilities in \eqref{eq1} is a
consequence of the usual Born rule along with the assumption that $M=0$ and
$M=1$ are equally probable.  Another framework can be used to discuss the
values of $S_x$ rather than $S_z$.  It is possible to incorporate all four
values of $M$, $0, 1, \bar 0, \bar 1$ --- note that these are four mutually
exclusive possibilities --- at the earlier time in either the $S_x$ or $S_z$
(at the later time) framework, but the $S_x$ and $S_z$ frameworks are
incompatible, since it makes no sense to simultaneously assign values of $S_x$
and $S_z$ to a spin-half particle.

	\subsection{Maximum capacity}
\label{sct4b}

	Consider a carrier of information described quantum mechanically by a
Hilbert space with a finite dimension $d$.  Then $d$ is the sum of the
dimensions of the different subspaces corresponding to the projectors making up
a decomposition of the identity, as is obvious by taking the trace of both
sides of \eqref{eq2}.  Consequently, there can be at most $d$ projectors in
such a decomposition of the identity, and the corresponding sample space
contains at most $d$ mutually exclusive possibilities.  For this reason, the
maximum amount of information that can be carried by this system is the same as
that of a classical object with $d$ different states, namely $\log_2 d$ bits.
In particular, for $d=2$, a single qubit, the maximum capacity is 1 bit; for
$d=2^n$, $n$ qubits, it is $n$ bits. Thus a quantum system can contain at most
one bit of information per qubit.

	One might suppose that this limit could be exceeded by using more than
one framework.  However, that would contradict the single framework rule: one
could not find a single sample space or a corresponding random variable
(Hermitian operator) to represent the additional information.  To better
understand what is involved in using different frameworks one needs to look at
various examples, which is what we shall do in 
the following sections. 
%%Secs.~\ref{sct5} and \ref{sct7}.

	\subsection{Location of information}
\label{sct4c}

	We looked at various examples in Sec.~\ref{sct2} where classical
information can be assigned a location in the sense that a carrier of
information (newspaper, colored slip of paper) can be assigned a location.  It
seems plausible that the same principle applies to the quantum case: if
a spin-half particle is in a particular location and its spin is appropriately
correlated with some initial message, then the information is located
wherever the particle is located.  Quantum particles cannot be assigned a
precise location, but neither can newspapers, so this is not a serious
difficulty.  Of greater concern is the fact that quantum theory allows the use
of different incompatible frameworks for describing quantum entities, and, as
we shall see, the presence or absence of information in some carrier of
information can depend upon the framework used to describe it.

	Let us begin with the simple case of a spin-half particle prepared by a
polarizer in the state $S_z=+1/2$ and traveling through a field-free region, as
in Sec.~\ref{sct3a}.  If we use the $S_z$ framework, this component of spin is
correlated with the switch setting $M$ on the polarizer, and the particle can
be said to contain or carry the information as to whether $M$ was 0 or 1.  If
instead one uses the $S_x$ framework, the joint probability of $M=0$ or 1 and
$S_x$ can be calculated using the Born rule, and one finds that they are
statistically independent, so in this framework there is no information about
the two switch settings. (For present purposes we are ignoring $M=\bar 0$ or
$\bar 1$.)  That the information should be present in one framework and not
another is not too surprising given the classical analogy (the second model)
introduced in Sec.~\ref{sct3b}: If the polarizer stores information in the $z$
component of angular momentum while leaving the $x$ and $y$ components
undefined or random, it is not surprising that the information is later found
in $S_z$ and not in $S_x$.  Thus in studying quantum information it is
important to look at a variety of frameworks.

	Consider a situation in which two spin-half particles are used to carry
two bits of information.  As long as each particle is polarized separately, the
situation is straightforward.  Suppose that polarizer $A$ imparts to particle
$a$ a polarization of $S_z=\pm 1/2$ and polarizer $B$ imparts to particle $b$ a
polarization of $S_x=\pm 1/2$.  Then one can exhibit perfect correlation
between the initial switch settings of the polarizers and the spin states of
the particles by using for the latter a framework consisting of the orthonormal
basis (with kets in the order $|a,b\rgl$) of:
\begin{equation}
  |0,+\rgl,\hquad |0,-\rgl,\hquad |1,+\rgl,\hquad |1,-\rgl,
\label{eq3}
\end{equation}
where $|0\rgl$ and $|1\rgl$ are the states with $S_z=+1/2$ and $-1/2$,
respectively, and 
\begin{equation}
  |+\rgl = \bigl(|0\rgl+ |1\rgl\bigr)/\st,\quad 
  |-\rgl = \bigl(|0\rgl- |1\rgl\bigr)/\st,
\label{eq4}
\end{equation}
are states with $S_x=+1/2$ and $-1/2$, with a particular choice of overall
phase. For this framework, information about $A$ will be located wherever
particle $a$ is located, information about $B$ will be at the location of
particle $b$, and the combined information can be said to be in the union of
the two regions of space occupied by the two particles.

	This is entirely analogous to the corresponding classical case and
would scarcely be worth mentioning except that it provides a contrast
with a very different way in which two bits of information can be encoded in
two spin-half particles by using the four Bell states
\begin{gather}
   |B_{00}\rgl =\bigl(|00\rgl + |11\rgl\bigr)/\st,\quad  
   |B_{01}\rgl =\bigl(|01\rgl + |10\rgl\bigr)/\st,  
\notag\\
  |B_{10}\rgl =\bigl(|00\rgl - |11\rgl\bigr)/\st,\quad
  |B_{11}\rgl =\bigl(|01\rgl - |10\rgl\bigr)/\st.
\label{eq5}
\end{gather}
As these form an orthonormal basis of the two spin system, they can be used as
a framework of mutually exclusive possibilities.  One could imagine that they
are produced by a suitable apparatus, with each state corresponding to one of
four settings of a switch on the apparatus.  Then the information about the
switch setting is located in the union of the regions occupied by the two
particles. However, in contrast to the previous case, it is not possible to say
that any part of this information is located in either particle by itself.

	To begin with, each of the Bell states is a delocalized entity of a
type which possesses no classical analog: it is impossible to ascribe
properties to the individual spins when they are in such an entangled state.
The reason is that a projector representing some property of particle $a$ is
necessarily of the form $\hf(I_a + S_{aw})$, where the subscript $a$ labels the
particle, $w$ indicates some direction in space, and $I$ is the identity
operator on the Hilbert space.  Such a projector does not commute with a
projector $|B_{jk}\rgl\lgl B_{jk}|$ representing one of the Bell states.  To
suppose that a quantum system can simultaneously possess properties
corresponding to two non-commuting projectors is to abandon standard quantum
mechanics in favor of a theory of hidden variables. And it violates the
single-framework rule.

	Even if a system has been prepared in one of the four Bell states by
some apparatus, there is no reason why at a later time we must describe it
using the Bell states as an orthonormal basis.  We could, instead, adopt a
basis in which the individual particles $a$ and $b$ are assigned specific
properties, for example, $S_{az}$ and $S_{bz}$.  If one works out the
corresponding probabilities (which are the same as if $S_{az}$ and $S_{bz}$
were measured), one will find that in this framework the difference between a
preparation which produced $|B_{00}\rgl$ or $|B_{10}\rgl$, on the one hand, and
$|B_{01}\rgl$ or $|B_{11}\rgl$ on the other, is present in the correlation
between the value of $S_{az}$ and $S_{bz}$, but not in the values of $S_{az}$
by itself or $S_{bz}$ by itself.  Other frameworks can be used to extract other
pieces of information, but in no case is any part of the information located in
one of the particles by itself. The situation is thus analogous to the example
of classical information encoded in two slips of paper of the same or of
different colors, Sec.~\ref{sct2}, and thus not available in either slip by
itself.

	\section{Dense Coding}
\label{sct5}

	\subsection{Circuit and unitary time development}
\label{sct5a}

	As noted in Sec.~\ref{sct1}, the phenomenon of dense (or superdense)
coding has been interpreted as indicating that a single qubit can somehow
contain two bits of information. If true, this would contradict the maximum
capacity result of Sec.~\ref{sct4b}, and both for this reason and also because it
provides an instructive example of using frameworks to locate information,
dense coding is worth examining in some detail.

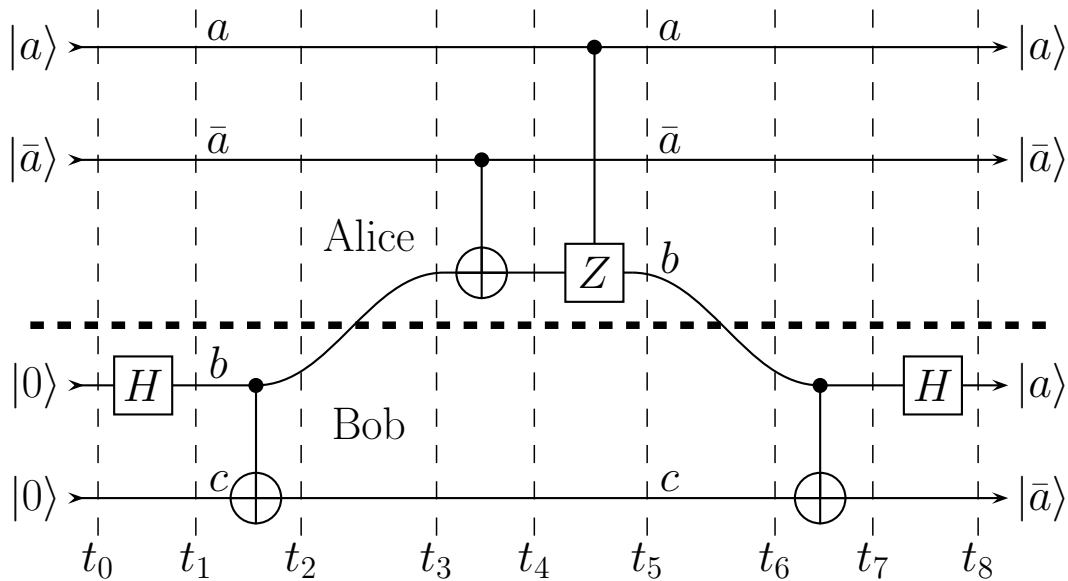
\begin{figure}
%r/figures/fg112.ptx   -*-tex-*- 11/12/01
% 11/12/01 Circuit for superdense coding
% Full size: Use first \begin{pspicture}, \Large, %unit=0.5
% Half size: Use second \begin{pspicture}, %\Large, unit=0.5
$$
	%PSPicture
\begin{pspicture}(-6.3,-1.2)(8.0,6.6) %Full size [(x,y) in cm]
%\begin{pspicture}(-3.2,-0.6)(4.0,3.3) %Half size
		%Grid
\newpsobject{showgrid}{psgrid}{subgriddiv=1,griddots=10,gridlabels=6pt}
%\showgrid
		% Parameters
\def\lwd{0.050} %Linewidth default
\def\rcp{0.35} %Radius of CNOT symbol, and its negative
\def\rdot{0.10} \def\rodot{0.15} %Radius of closed and open dot
\def\bxp{0.40} %Half edge of square box
%\def\vtl{linestyle=dashed,dash=7pt 7pt,linewidth=0.02} %vertical dashed lines
		% Defaults
\Large
%\psset{%unit=0.5cm,
%arrowsize=0.20,linewidth=0.050}
		% Object definitions
\def\cnot{\pscircle(0,0){\rcp}\psline(-\rcp,0)(\rcp,0)\psline(0,-\rcp)(0,\rcp)}
\def\dot{\pscircle*(0,0){\rdot}}
\def\squ{%  Square box representing gate
\psframe[fillcolor=white,fillstyle=solid](-\bxp,-\bxp)(\bxp,\bxp)}% END squ
\def\time#1#2{\psline[linestyle=dashed,dash=0.25 0.25,linewidth=0.02]%
(#2,6.5)(#2,-0.5)\rput[t](#2,-0.6){$t_{#1}$}}%END time
		% Diagram
	%Lines
\psline{>->}(-5.5,0.0)(7.0,0.0)
\psline{>->}(-5.5,4.5)(7.0,4.5)
\psline{>->}(-5.5,6.0)(7.0,6.0)
\psline{>-}(-5.5,1.5)(-3.0,1.5)
\psbezier(-3.0,1.5)(-2.0,1.5)(-1.5,3.0)(-0.5,3.0)
\psline(-0.5,3.0)(2.0,3.0)
\psbezier(2.0,3.0)(3.0,3.0)(3.5,1.5)(4.5,1.5)
\psline{->}(4.5,1.5)(7.0,1.5)
\psline[linestyle=dashed,dash=0.18 0.18,linewidth=0.1](-6.0,2.3)(7.5,2.3)
	%Time lines
\time{0}{-5.1}\time{1}{-3.8}\time{2}{-2.4}\time{3}{-0.6}
\time{4}{0.7}\time{5}{2.2}\time{6}{3.9}\time{7}{5.2}\time{8}{6.6}
	%Cnots
\rput(-3.0,0.0){\cnot} \rput(0.0,3.0){\cnot} \rput(4.5,0.0){\cnot}
\rput(-3.0,1.5){\dot} \rput(0.0,4.5){\dot} \rput(4.5,1.5){\dot}
\psline(-3.0,0.0)(-3.0,1.5) \psline(0.0,3.0)(0.0,4.5)
\psline(4.5,0.0)(4.5,1.5)
	%Z gate
\rput(1.5,6.0){\dot}
\psline(1.5,6.0)(1.5,3.0)
\rput(1.5,3.0){\squ}
\rput(1.5,3.0){$Z$}
	%H gates
\rput(-4.5,1.5){\squ}
\rput(-4.5,1.5){$H$}
\rput(6.0,1.5){\squ}
\rput(6.0,1.5){$H$}
	% Text
\rput[r](-5.6,0.0){$|0\rgl$}
\rput[r](-5.6,1.5){$|0\rgl$}
\rput[r](-5.6,4.5){$|\bar a\rgl$}
\rput[r](-5.6,6.0){$| a\rgl$}
\rput[l](7.1,0.0){$|\bar a\rgl$}
\rput[l](7.1,1.5){$| a\rgl$}
\rput[l](7.1,4.5){$|\bar a\rgl$}
\rput[l](7.1,6.0){$| a\rgl$}
\rput[b](-3.5,6.1){$a$}
\rput[b](-3.5,4.6){$\bar a$}
\rput[b](-3.5,1.6){$b$}
\rput[b](-3.5,0.1){$c$}
\rput[b](2.5,6.1){$a$}
\rput[b](2.5,4.6){$\bar a$}
\rput[b](2.5,3.0){$b$}
\rput[b](2.5,0.1){$c$}
\rput(-1.5,3.5){\psframebox*{Alice}}
\rput(-1.5,1.0){\psframebox*{Bob}}
\end{pspicture}
$$
\caption{%
Quantum circuit for dense coding.}
\label{fg1}
\end{figure}

	A quantum circuit corresponding to this process is shown in
Fig.~\ref{fg1}, with the usual convention that time increases from left to
right. Each of the two qubits $a$ and $\bar a$ in Alice's laboratory, above the
dashed line, is initially in a state $|0\rgl$ or $|1\rgl$, and together they
constitute an information source with four possible messages.  In Bob's
laboratory, below the dashed line, two qubits $b$ and $c$, initially both in
the $|0\rgl$ state are entangled by applying a Hadamard gate $H$ to $|b=0\rgl$,
producing $(|0\rgl+|1\rgl)/\st$, followed by a controlled-not in which the $c$
qubit is flipped from $|0\rgl$ to $|1\rgl$, or vice versa, if and only if the
$b$ qubit is in the state $|1\rgl$, where the notation follows \cite{NiCh00}.
These two unitary transformations place the two qubits $b$ and $c$ in the Bell
state $|B_{00}\rgl$, see \eqref{eq5}.  At this point the $b$ qubit is
transported to Alice's laboratory and the information in the $a$ and $\bar a$
qubits is written into it using a controlled-not and a controlled-$Z$ gate.
The latter multiplies the total state by $-1$ if and only if both $a$ and $b$
are in the state $|1\rgl$. After this the $b$ qubit is returned to Bob's
laboratory where it and the $c$ qubit undergo the inverse of the operation that
initially entangled them. The overall unitary time evolution results in the $b$
qubit being in the same state, $|0\rgl$ or $|1\rgl$, as the $a$ qubit, and $c$
in the same state as $\bar a$, as indicated in the figure. (Note that if $a$ is
in a superposition of $|0\rgl$ and $|1\rgl$, this same state will \emph{not}
emerge at the final time at either of the points labeled $|a\rgl$ in the
diagram, and the same is true of $\bar a$.)
	The term ``dense coding'' refers to the idea, which at first glance
seems intuitively plausible, that the single qubit $b$ on its way back to Bob's
laboratory contains the information that was initially in both $a$ and $\bar
a$, which is to say two bits of information, and it is this idea that we shall
investigate using various frameworks.  

	\subsection{Different frameworks}
\label{sct5b}

Let us begin with the framework $\FS_1$ corresponding to the unitary time
development of the initial states discussed above.  One then finds that during
the time interval from $t_5$ to $t_6$ in Fig.~\ref{fg1} the qubits $b$ and $c$
are in one of the entangled Bell states defined in \eqref{eq5}; which state
depends on the values of $a$ and $\bar a$.  But in this situation, as discussed
in the Sec.~\ref{sct4c}, the information cannot be said to be present in qubit
$b$, instead it is in a sort of quantum correlation between $b$ and $c$.
Consequently, if we use $\FS_1$ there is no violation of the maximum capacity
result of Sec.~\ref{sct4b}, because 2 bits of information are carried by a
Hilbert space of dimension $2\times 2=4$ representing both qubits.

	The reader may find this result, even though formally correct, to be
intuitively troubling in that it seems as if the information which at times
before $t_3$ was entirely in Alice's laboratory has somehow managed to
``jump'', so that at time $t_5$ it is at least partly in Bob's laboratory in
qubit $b$, despite the fact that this qubit has not interacted with either $a$
or $\bar a$.  Have we uncovered some mysterious long-range interaction which
can influence $b$ despite the fact that it is far away from Alice's laboratory?

	To see that there is nothing particularly odd about information
suddenly appearing in this manner in a delocalized correlation with a distant
object, it is useful to analyze the circuit in Fig.~\ref{fg1} using a different
framework $\FS_2$ in which we suppose that at $t_2$ and all later times each
qubit is described using the $S_z$ or computational basis, $|0\rgl$ or
$|1\rgl$, which for this discussion we shall denote by 0 or 1 without using the
ket symbol.  (The state of $b$ at $t_8$ is an exception, see below.)  Given the
initial states shown in the figure, it follows from the Born rule that in this
framework at time $t_2$ either $b=c =0$ or $b=c =1$, with equal probability.
The same is true at $t_3$, but between $t_3$ and $t_4$ the $b$ qubit will be
flipped from $0$ to $1$ or from $1$ to $0$ if $\bar a=1$, or remain the same if
$\bar a = 0$.  As a consequence, at $t_4$ the information initially present in
$\bar a$ will be present in the \emph{correlation} between $b$ and $c$: if
$\bar a=0$, then $b=c$, and if $\bar a =1$, then $b\neq c$.  Notice that this
information is \emph{not} present in either $b$ or $c$ separately, for at $t_4$
each is 0 or 1 with a probability of 1/2, the same as at $t_3$.  Once again we
have a situation in which information ``suddenly'' (between $t_3$ and $t_4$)
appears as a correlation with a distant object which has taken no part in any
interaction.  However, the situation is precisely parallel with the classical
analogy discussed in Sec.~\ref{sct2}, in which Alice and Bob share two slips of
paper which are both red or both green, and there are obviously no nonlocal
influences or other sorts of magic.  And if such an information ``jump'' is
unproblematic in the classical case, and thus in framework $\FS_2$, there seems
to be no reason to be concerned about it in framework $\FS_1$.

	Now let us continue the description using framework $\FS_2$ to times
later than $t_4$ in Fig.~\ref{fg1}.  Between $t_4$ and $t_5$ the $b$ qubit is
left unchanged, for even if $a=1$ the only effect of this is to change the sign
of \hbox{$|b=1\rgl$}, but this has no effect upon the projector $|1\rgl\lgl
1|$.  (This is perhaps clearer if one employs the histories formalism, see the
next paragraph.)  Nor is there, obviously, any change between $t_5$ and $t_6$.
The next controlled-not gate has the effect that $c$ is equal to 0 or 1 at
$t_7$ depending upon whether $b=c$ or $b\neq c$ at
$t_6$. Consequently, the information originally contained in $\bar a$, which
during the time interval from $t_4$ to $t_6$ is present as a correlation
between $b$ and $c$, emerges at the end in $c$, as indicated in the
figure.  The final Hadamard gate changes qubit $b$ from $|0\rgl$ to $|+\rgl$ or
$|1\rgl$ to $|-\rgl$ as the case may be; one cannot in $\FS_2$ assign $|0\rgl$
or $|1\rgl$ at $t_8$ to $b$ without violating consistency (decoherence)
conditions \cite{nt3}.

	The somewhat informal discussion of framework $\FS_2$ in the previous
paragraphs can be given a formal justification using the consistent histories
formalism, where the notation is that of \cite{Grf02}.  Consider a
consistent family with four initial states
\begin{equation}
 |\Psi_0^{a\bar a}\rgl =|a,\bar a,b=0,c=0\rgl,  
\label{eq6}
\end{equation} 
where $a$ and $\bar a$ are either 0 or 1, 
and a set of histories of the form
\begin{equation}
  [\Psi_0^{a\bar a}] \od F_2 \od F_4 \od F_7,
\label{eq7}
\end{equation} 
where the subscripts refer to the times shown in Fig.~\ref{fg1}.  (One could
include additional events at additional times, but this selection suffices for
displaying the essential ideas.)  In \eqref{eq7} and below we use the
abbreviation $[\Phi]$ for the projector $|\Phi\rgl\lgl\Phi|$.  For each
possible initial state there are two histories
\begin{equation}
\newcommand{\amp}{&\hskip -.7em\relax } 
\newcommand{\oda}{\amp\od\amp }
\bem
  {}[\Psi_0^{a\bar a}] \oda [a,\bar a,0,0]_2\oda [a,\bar a,\bar a,0]_4
  \oda[a,\bar a,\bar a,\bar a]_7,
\\[0.65 ex]
  {}[\Psi_0^{a\bar a}] \oda [a,\bar a,1,1]_2\oda [a,\bar a,1-\bar a,1]_4
  \oda[a,\bar a,1-\bar a,\bar a]_7
\enm
\label{eq8}
\end{equation} 
with non-zero probabilities.  Here subscripts have been added to the projectors
to identify the time.  For a given $a$ and $\bar a$ the two non-zero
probabilities are equal to each other; they can be calculated using the methods
discussed in \cite{Grf96} or in Chs.~10 and 11 of \cite{Grf02}.  From these
one can see that at $t_4$ information corresponding to the value of $\bar a$
is, indeed, present in the correlation between the last two qubits $b$ and $c$,
but not in either of these separately, because one does not know which of the
histories in
\eqref{eq8} is the one that actually occurs.

	Finally, let us consider a third framework $\FS_3$ which is similar to
$\FS_2$ except that for times in the interval $t_2 \leq t \leq t_6$ we use the
$S_x$ basis for qubits $b$ and $c$, i.e., the states $|+\rgl$ and $|-\rgl$
defined in \eqref{eq4}, while retaining $|0\rgl$ and $|1\rgl$ for $a$ and $\bar
a$.  Since $b$ and $c$ evolve unitarily to
\begin{equation}
  |B_{00}\rgl = \bigl( |00\rgl + |11\rgl\bigr)/\st = (|++\rgl + |--\rgl)/\st
\label{eq9}
\end{equation} 
at time $t_2$, by thinking of this state as a pre-probability \cite{nt0}, or by
invoking the Born rule, one sees that in framework $\FS_2$, $b=c = +$ and $b=c
= -$ occur with equal probability.  The controlled-not between $t_3$ and $t_4$
has no effect upon $b$ in this basis, whereas the controlled-Z between $t_4$
and $t_5$ interchanges $+$ and $-$ if $a=1$, or leaves $b$ the same if $a=0$.
Thus in the time interval between $t_5$ and $t_6$ the information originally
present in $a$ is found in the correlation $b=c =+$ or $-$ for $a=0$, and
$b\neq c$ for $a=1$. The situation is thus analogous to what we found using
$\FS_2$, but with the roles of $a$ and $\bar a$ interchanged. The
controlled-not between $t_6$ and $t_7$ maps $b=c$ into $b=+$, and $b\neq c$
into $b=-$, while the final Hadamard gate maps these to $b=0$ and $b=1$,
respectively, so that at the end $b$ is identical to $a$.  On the other hand,
$c$ continues to have the same value, $+$ or $-$, at $t_7$ and $t_8$ as it had
earlier; ascribing a value of 0 or 1 to it at these times would violate the
consistency conditions \cite{nt3}.

	In terms of the location of information, the $\FS_3$ framework leads to
a conclusion parallel to the one obtained using $\FS_2$: during the time
interval from $t_5$ to $t_6$ the information about $a$ is located in a
correlation between $b$ and $c$ in a manner which is, once again, quite
analogous to the classical situation considered in Sec.~\ref{sct2}.  In
particular one cannot say that the information is present in $b$ by itself
any more than that it is in $c$ by itself. 

	\subsection{Discussion}
\label{sct5c}

	It should be noted that the three frameworks $\FS_1$, $\FS_2$, and
$\FS_3$ are incompatible, as is immediately obvious from the fact that they use
noncommuting projectors to describe the $b$ and $c$ qubits at $t_2$ and at each
later time.  Thus the descriptions provided by these frameworks cannot be
combined into a single quantum description; this would be just as meaningless
as assigning values to both $S_x$ and $S_z$ for a spin half particle.
Nonetheless, they provide qualitatively similar answers to the question of the
location of the information transmitted from Alice to Bob: it is not to be
found in the $b$ qubit alone during the time interval between $t_5$ and $t_6$,
but instead in some sort of correlation between $b$ and $c$.  In $\FS_1$ this
``nonlocality'' arises from the delocalized nature of entangled quantum states,
which has no classical analog, whereas in $\FS_2$ and $\FS_3$ it is rather more
like the classical nonlocality discussed in Sec.~\ref{sct2}.  In addition, in
$\FS_2$ and $\FS_3$, half of the information --- a different half in the two
cases --- has completely disappeared by the time $t=t_5$, and does not
re-emerge. That information should be ``invisible'' in certain frameworks is
not by itself too surprising, since there is a classical analog,
Sec.~\ref{sct3b}: the $x$ component of angular momentum of a spinning object
reveals no information about the $z$ component.  However, the inability to
combine these incompatible frameworks into a single description is very much a
quantum effect.

	Note that in no case is the maximum capacity limit of Sec.~\ref{sct4b}
exceeded, because with each of these frameworks the information is carried
jointly by $b$ and $c$, and not by $b$ alone.  One might worry that there could
be some other framework in which all of the information is found in $b$, but
one can easily show that this is impossible by calculating the reduced density
matrix for $b$ at time $t_5$ using unitary time development from any of the
initial states, and verifying that it is 1/2 times the identity operator
whatever may be the initial choice of $a$ and $\bar a$.  This density matrix
functions as a pre-probability \cite{nt4}, i.e., it can be used to assign
probabilities for any basis in the Hilbert space of $b$, and the corresponding
probabilities are thus independent of the values of $a$ and $\bar a$.

	Difficulties in understanding dense coding have led to the suggestion
\cite{nt5} that somehow some of the information present in $a$ and $\bar a$
travels backwards in time in order to reach Bob's laboratory.  However, it is
easy to show that given the initial states shown in Fig.~\ref{fg1}, at any time
$t\leq t_3$ any random variable (i.e., any Hermitian operator) on the Hilbert
space of the two qubits $b$ and $c$ is uncorrelated with $a$ and $\bar a$.
Thus the information does not travel backwards in time, except, perhaps, in
some metaphorical sense, and then one would expect this to be equally true in
the corresponding classical case discussed in Sec.~\ref{sct2}.

	Nonetheless there \emph{is} something very odd, that is to say
nonclassical, about dense coding.  In the classical analog,
while it is possible to encode information in correlations between two carriers
of information, each with a maximum capacity of 1 bit, one can insert at most a
single bit of information into this system by using interactions affecting
\emph{only one} of the two carriers.  In the quantum case the situation is
different, for starting with a single Bell state one can generate any of the
other three by applying an appropriate unitary interaction to just \emph{one}
of the qubits, as noted in \cite{BeWi92}.  In this respect
quantum theory is very different from classical physics, and one has a genuine
quantum mystery.  But it is a mystery fully compatible with the information
bound in Sec.~\ref{sct4b}, and with the absence of mysterious nonlocal
influences which can transfer information instantly between separated (thus
noninteracting) systems.

	\section{Structures and Channels}
\label{sct6}

	\subsection{Structures}
\label{sct6a} 

	Thus far we have discussed information in the quantum context in terms
of frameworks.  As long as a single framework is employed for describing a
quantum system, the usual rules which govern classical probabilities and
ordinary (classical) information theory apply, so in this sense there is
nothing new.   However, in quantum theory, in contrast to
classical physics, one has the possibility of using many different incompatible
frameworks, and each framework has a distinct probabilistic structure.  This
\emph{framework dependence} is something with no classical analog, and
gives rise to problems of a sort not encountered in other uses of information
theory.  Examples of framework dependence have already come up in the
discussion of dense coding in Sec.~\ref{sct5}.

	For discussions of framework dependence it is useful to introduce the
concept of a \emph{structure}: a collection of frameworks, typically mutually
incompatible, along with a common algorithm for assigning probabilities in the
different frameworks.  This is not a precise definition, for the concept is
still under development, but a few examples will serve to indicate what we have
in mind.  A particular density matrix can be used to assign probabilities to
the elements of any orthonormal basis of the Hilbert space on which it is
defined, and in this case the bases are the different frameworks that
constitute the structure, while the density matrix provides the probability
algorithm.  As another example, suppose Eve is contemplating how best to
extract information about the common key which Alice and Bob are in the process
of establishing using the BB84 quantum cryptographic protocol \cite{BeBr84}.
She has to design an apparatus to intercept and in some sense copy a qubit on
the way from Alice to Bob, and in working out the design needs to compute what
will happen using either an $S_x$ or $S_z$ framework for the initial state of
the qubit, as well as various alternatives at later times.  The dynamical
properties of the eavesdropping device provide the algorithm for calculating
probabilities for the different scenarios, each corresponding to some
particular framework.

	\subsection{Channels}
\label{sct6b}

	Perhaps the simplest example of a structure involving histories of a
quantum system, thus properties at more than one time, is a \emph{quantum
channel}.  Typically, one is interested in some basis for the Hilbert space
representing the input to the channel, which is tensored to a second Hilbert
space representing the environment, and then another (possibly the same) basis
for the first space at a later time.  A particular framework corresponds to a
particular choice of the initial and final bases, the structure consists of all
frameworks of this sort, and unitary time evolution of the total system
(channel plus environment) along with the Born rule is used to calculate
probabilities in each framework.

	Let $\AS$ be the input of the channel, $\ES$ the environment, and
$\HS=\AS\ot\ES$ the total Hilbert space.  Suppose that at $t_0$ the environment
is always in a fixed state $|e_0\rgl$, while the initial state $|a\rgl$ of the
channel is one of a set of states $\{|p^m\rgl\}$, $m=1,2,\ldots$, which form an
orthonormal basis of $\AS$.  At a time $t_1 > t_0$, define
\begin{equation}
    |\Psi_1^m\rgl = T(t_1,t_0)\bigl( |p^m\rgl|e_0\rgl\bigr),
\label{eq10}
\end{equation}
where $T(t',t)$ is the unitary time development operator, equal to 
$\exp[-i(t'-t)H/\hbar]$ for a time-independent Hamiltonian $H$. 

	Next consider a framework consisting of a family of histories, each of
which begins with some initial state $|p^m\rgl|e_0\rgl$ at $t_0$, and ends with
one of the events associated with a decomposition of the identity
\begin{equation}
    I= \bar P + \sum_m P^m
\label{eq11}
\end{equation}
in orthogonal projectors, where $\bar P$ might be zero, and the 
$\{P^m\}$ have the property that
\begin{equation}
    P^m|\Psi_1^{m'}\rgl = \dl_{mm'}|\Psi_1^{m}\rgl.
\label{eq12}
\end{equation}
As a consequence, information about the basis states $\{|p^m\rgl\}$ at $t_0$ is
at time $t_1$ contained in the collection of projectors $\{P^m\}$. 

	Suppose that the $\{P^m\}$ have been chosen in such a way that for some
factorization $\BS\ot\FS$ of $\HS$, which could but need not be the same as
$\AS\ot\ES$, it is the case that each projector $P^m$ is \emph{on} $\BS$ in the
sense that it is an operator of the form $P^m\ot I$ (which means $\bar P$ is
also of this form). Then we can say that the information about $\{|p^m\rgl\}$
is \emph{in} $\BS$ (or ``on'' or ``carried by'' $\BS$).  Typically $\BS$ will
refer to some subsystem of the total system described by $\HS$, and if this
subsystem is located in a particular region of space, then the information can
be said to be located in that region.

	One can always find projectors such that \eqref{eq12} is satisfied, for
example, 
\begin{equation}
    P^m = |\Psi_1^m\rgl\lgl \Psi_1^m|. 
\label{eq13}
\end{equation}
However, this choice is so precise that $\BS$ will be equal to $\HS$, and to
find more interesting cases in terms of location of information one needs to
choose coarser projectors.  Some examples will be found in Sec.~\ref{sct7b}, or
the reader may wish to think about the case in which the time transformation is
trivial, $T(t',t)=I$,  and $\BS=\AS$.

	In the case of a quantum channel one is usually not interested in
just a single orthonormal basis of $\AS$ at $t_0$, but instead all possible
orthonormal bases.  Let $\{|p^m_\lm\rgl,m=1,2,\ldots\}$ be the collection of
vectors forming the orthonormal basis $\lm$.  Here $\lm$ can be any convenient
label or set of parameters used to label the bases, but it should be chosen so
that it does not distinguish bases which differ only through relabeling the
basis vectors, or multiplying them by phase factors: in other words, it should
label distinct decompositions of the identity into pure states.  For example,
for a spin-half particle or a single qubit, with $|0\rgl$ and $|1\rgl$ the eigenstates of $S_z$ with
eigenvalues $+1/2$ and $-1/2$, define
\begin{equation}
    |p^m_\lm\rgl = \al^m_\lm |0\rgl + \bt^m_\lm |1\rgl, 
\label{eq14}
\end{equation}
with $m=0$ or 1, where
\begin{equation}
    \al^0_\lm = \frac{1}{\sqrt{1+|\lm|^2}},\quad 
   \bt^0_\lm = \frac{\lm}{\sqrt{1+|\lm|^2}},\quad 
    \al^1_\lm = \frac{\lm^*}{\sqrt{1+|\lm|^2}},\quad 
   \bt^1_\lm = \frac{-1}{\sqrt{1+|\lm|^2}},
\label{eq15}
\end{equation}
for $\lm$ a complex number satisfying $|\lm| < 1$, or $\lm = e^{i\phi}$
with $0\leq \phi < \pi$. In particular, $\lm=0$ is the $S_z$ or computational
basis, $\lm=1$ the $S_x$ basis, and $\lm=i$ the $S_y$ basis in the familiar
Bloch sphere picture.

	Let us in addition allow many different times: $t_0 < t_1 < t_2 <
\cdots < t_f$. For each basis $\lm$ and each time $t_j$ choose a decomposition
of the identity
\begin{equation}
    I=\bar P_{\lm j} + \sum_m P^m_{\lm j}
\label{eq16}
\end{equation}
such that 
\begin{equation}
    P^m_{\lm j}|\Psi_{\lm j}^{m'}\rgl = \dl_{mm'}|\Psi_{\lm j}^{m}\rgl,
\label{eq17}
\end{equation}
where
\begin{equation}
    |\Psi^m_{\lm j}\rgl = T(t_j,t_0)\bigl( |p^m_\lm\rgl |e_0\rgl \bigr).
\label{eq18}
\end{equation}

	Next suppose that at time $t_j$ there is a factorization
$\BS_j\ot\FS_j$ of $\HS$ such that for all $\lm$ and every $m$, $P^m_{\lm j}$ is
\emph{on} $\BS_j$. Then since for any $\lm$ the information about
$\{|p^m_\lm\rgl\}$ is in $\BS_j$, it seems reasonable to say that the channel
itself is \emph{in} (or ``on'' or ``carried by'') $\BS_j$ at time $t_j$.  Let
us adopt this as a tentative definition that deserves further exploration to
see if it is reasonable and useful.  That exploration begins in
Sec.~\ref{sct7}.

	Note that for a fixed $\lm$ the set of histories
\begin{equation}
    [p^m_\lm]\od P^m_{\lm 1}\od P^m_{\lm 2}\od \cdots \od P^m_{\lm f},
\label{eq19}
\end{equation}
with $m=1,2,\ldots$, form the support of a consistent family or framework
$\FS_\lm$ in the notation of \cite{Grf02}.  Consistency is easily established
using the method of chain kets \cite{nt6}.  Note that $\FS_\lm$ and
$\FS_{\lm'}$ are incompatible for $\lm\neq\lm'$, so it is not correct to think
of information about $\{|p^m_\lm\rgl\}$ and $\{|p^m_{\lm'}\rgl\}$ as
simultaneously present in $\BS_j$ even when the channel itself, viewed as a
structure, can be said to be present in $\BS_j$.

	\section{Teleportation Circuit}
\label{sct7}

	\subsection{Circuit and unitary time development}
\label{sct7a}

\begin{figure}
$$
%r/figures/fg113.ptx   -*-tex-*- 
% 11/12/01 Circuit for teleportation
	%PSPicture
\begin{pspicture}(-6.3,-1.2)(8.0,5.0) %(x,y) lower left, upper right in cm
		%Grid
\newpsobject{showgrid}{psgrid}{subgriddiv=1,griddots=10,gridlabels=6pt}
%\showgrid
		% Parameters
\def\lwd{0.050} %Linewidth default
\def\rcp{0.35} %Radius of CNOT symbol, and its negative
\def\rdot{0.10} \def\rodot{0.15} %Radius of closed and open dot
\def\bxp{0.40} %Half edge of square box
		% Defaults
%\Huge
\Large
%\psset{%unit=0.7cm,
%arrowsize=0.20,linewidth=\lwd}
		% Object definitions
\def\cnot{\pscircle(0,0){\rcp}\psline(-\rcp,0)(\rcp,0)\psline(0,-\rcp)(0,\rcp)}
\def\dot{\pscircle*(0,0){\rdot}}
\def\squ{%  Square box representing gate
\psframe[fillcolor=white,fillstyle=solid](-\bxp,-\bxp)(\bxp,\bxp)}% END squ
\def\time#1#2{\psline[linestyle=dashed,dash=7pt 7pt,linewidth=0.02]%
(#2,5.1)(#2,-0.5)\rput[t](#2,-0.6){$t_{#1}$}}%END time
		% Diagram
	%Lines
\psline{>->}(-5.5,0.0)(7.0,0.0)
\psline{>->}(-5.5,4.5)(7.0,4.5)
\psline{>-}(-5.5,1.5)(-3.0,1.5)
\psbezier(-3.0,1.5)(-2.0,1.5)(-1.5,3.0)(-0.5,3.0)
\psline{->}(-0.5,3.0)(7.0,3.0)
\psline[linestyle=dashed,dash=5pt 5pt,linewidth=0.1](-6.0,2.3)(3.0,2.3)
	%Time lines
\time{0}{-5.1}\time{1}{-3.8}\time{2}{-2.4}\time{3}{-0.6}
\time{4}{0.7}\time{5}{2.2}\time{6}{3.9}\time{7}{5.2}\time{8}{6.6}
	%Cnots
\rput(-3.0,0.0){\cnot} \rput(0.0,3.0){\cnot} \rput(4.5,0.0){\cnot}
\rput(-3.0,1.5){\dot} \rput(0.0,4.5){\dot} \rput(4.5,3.0){\dot}
\psline(-3.0,0.0)(-3.0,1.5) \psline(0.0,3.0)(0.0,4.5)
\psline(4.5,0.0)(4.5,3.0)
	%Z gate
\rput(6.0,4.5){\dot}
\psline(6.0,4.5)(6.0,0.0)
\rput(6.0,0.0){\squ}
\rput(6.0,0.0){$Z$}
	%H gates
\rput(-4.5,1.5){\squ} \rput(1.5,4.5){\squ}
\rput(-4.5,1.5){$H$} \rput(1.5,4.5){$H$}
		% Text
\rput[r](-5.6,0.0){$|0\rgl$}
\rput[r](-5.6,1.5){$|0\rgl$}
\rput[r](-5.6,4.5){$|\psi\rgl$}
\rput[l](7.1,0.0){$|\psi\rgl$}
\rput[l](7.1,3.0){$|+\rgl$}
\rput[l](7.1,4.5){$|+\rgl$}
\rput(-1.5,3.5){\psframebox*{Alice}}
\rput(-1.5,1.0){\psframebox*{Bob}}
\rput[b](-3.5,4.6){$a$}
\rput[b](-3.5,1.6){$b$}
\rput[b](-3.5,0.1){$c$}
\rput[b](3.0,4.6){$a$}
\rput[b](3.0,3.1){$b$}
\rput[b](3.0,0.1){$c$}
\end{pspicture}
$$
\caption{%
Quantum circuit for teleportation.}
\label{fg2}
\end{figure}
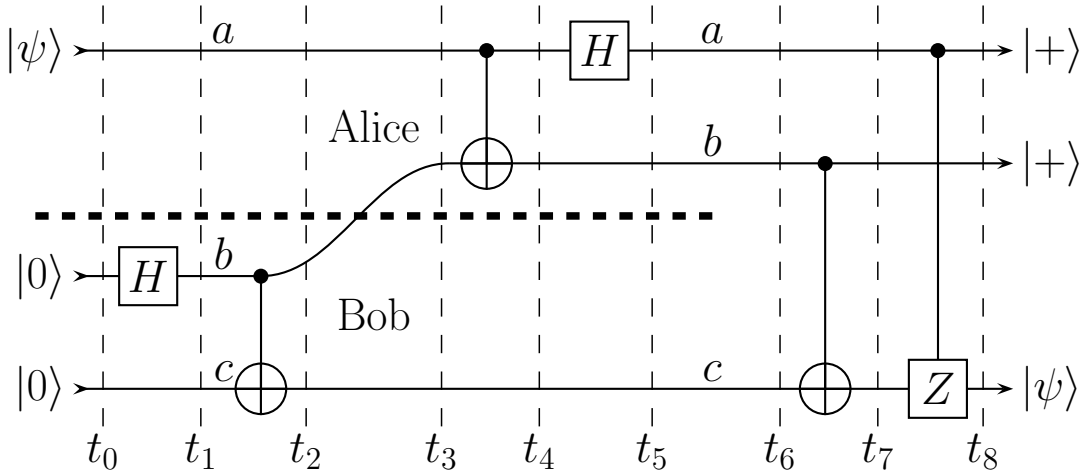

	The circuit shown in Fig.~\ref{fg2}, with labels chosen to facilitate
comparison with Fig.~\ref{fg1}, was first proposed (in a slightly different
notation) as a model for teleportation in \cite{BrBC98}; also see p.~187 of
\cite{NiCh00}.  The unknown state $|\psi\rgl$ to be teleported is initially in
qubit $a$, whereas $b$ and $c$ at time $t_2$ are in the Bell state
$|B_{00}\rgl$, see \eqref{eq5}.  If qubits $a$ and $b$ are in one of the Bell
states $|B_{jk}\rgl$ at $t_3$, the two gates which follow will put them in a
product state $|ab\rgl = |jk\rgl$.  Consequently, a measurement of both $a$ and
$b$ at time $t_5$ in the computational basis is equivalent to a measuring the
pair of qubits $a,b$ at $t_3$ in the Bell basis \cite{nt7}. In the standard
teleportation scenario Alice sends the results of such measurements over a
classical channel to Bob, who carries out appropriate unitary operations on
qubit $c$.  There are no measurements in the circuit in Fig.~\ref{fg2} and the
final unitary operations on $c$ are carried out ``automatically'' by the last
two gates. Thus this circuit does not actually represent teleportation, though
it can be viewed in a particular way using an appropriate structure,
Sec.~\ref{sct7d} below, which makes it essentially equivalent to teleportation.
However, it is also of interest as a nontrivial example of an ideal quantum
channel, and we shall begin by studying it from this point of view, in
particular addressing the question of the location of information in various
frameworks.

	The circuit in Fig.~\ref{fg2} constitutes a quantum channel in the
sense defined in Sec.~\ref{sct6b}, where qubits $b$ and $c$ in the state
$|e_0\rgl=|00\rgl$ at $t_0$ are the 
environment, while the channel input can be thought of as being in a state
\begin{equation}
    |a\rgl = \al|0\rgl +\bt|1\rgl, \quad |\al|^2 + |\bt|^2 =1.
\label{eq20}
\end{equation}
Unitary time development will then result in a state $|\Psi_k\rgl$ for the
total system at time $t_k$, with, in particular,
\begin{equation}
   |\Psi_2\rgl = (\al|0\rgl + \bt|1\rgl)\ot\bigl(|00\rgl + |11\rgl\bigr)/\st, 
\label{eq21}
\end{equation}
\begin{align}
   |\Psi_5\rgl &= (1/\st)\Bigl[ (\al|+\rgl + \bt|-\rgl)\ot|++\rgl 
  +(\al|+\rgl - \bt|-\rgl)\ot|--\rgl \Bigr]
\notag\\
   &= \hf\Bigl[|00\rgl\ot(\al|0\rgl + \bt|1\rgl) +
 |01\rgl\ot(\bt|0\rgl + \al|1\rgl) 
\notag\\
  &\hphantom{=}+|10\rgl\ot(\al|0\rgl - \bt|1\rgl) +
 |11\rgl\ot(-\bt|0\rgl + \al|1\rgl)\Bigr],
\label{eq22}
\end{align}
\begin{equation}
     |\Psi_8\rgl = |++\rgl\ot(\al|0\rgl + \bt|0\rgl). 
\label{eq23}
\end{equation}
Here $|+\rgl$ and $|-\rgl$ are the $S_x$ basis states defined in \eqref{eq4},
the qubits are in the order $|abc\rgl$, and the tensor product symbol $\ot$ has
been inserted at various points for clarity. 

	\subsection{Location of the channel}
\label{sct7b}

	Let us now consider a particular basis $\lm$ for the $a$ qubit at
$t_0$, with the basis states given by $|a\rgl= |p^m_\lm\rgl$, \eqref{eq14},
with coefficients defined in \eqref{eq15}.  Then it is easy to construct a
framework in which at $t_3$ the information about the initial states is found
in qubit $a$ using the criteria of Sec.~\ref{sct6b}.  In particular, the
decomposition of the identity \eqref{eq16} is of the form
\begin{equation}
    I = [p^0_\lm]\ot I\ot I +  [p^1_\lm]\ot I\ot I
\label{eq24}
\end{equation}
with projectors which are on qubit $a$, and one can check
that the conditions \eqref{eq17} are satisfied.  As this is true for any choice
of basis $\lm$, we can, using the
definition given in  Sec.~\ref{sct6b},  say that the channel itself is
\emph{in} qubit $a$ at $t_3$, and of course the same argument will work at
$t_1$ and $t_2$. This agrees with one's intuition in that up to time $t_3$ the
$a$ qubit has not interacted with the other qubits, so one would expect that
any information initially present in any basis, and thus the channel itself,
would continue to be present in qubit $a$.  Note, by the way, that for all
$t\leq t_3$ and for any choice of initial basis $\lm$, there is no information
about the initial state of $a$ in qubits $b$ or $c$ or the combination of the
two --- the situation is analogous to the case of dense coding, see the
discussion in Sec.~\ref{sct5} --- so once again there is no question of
information in some way traveling backwards in time through either $b$ or $c$
\cite{nt8}.

	Next let us consider the situation at time $t_5$.  First, it is easy to
show that except for special choices of $\lm$ the information about the initial
state is not present (or at least not completely present) in qubit $a$ by
itself, and therefore the quantum channel cannot be said to be present in
$a$. However, it is present in the \emph{pair} $a,b$.  This can be seen most
easily by noting that in the first expression for $|\Psi_5\rgl$ in
\eqref{eq22}, the $|+\rgl$ and $|-\rgl$ states for qubit $b$ occur in
conjunction with $\al|+\rgl + |\bt|-\rgl$ and $\al|+\rgl - |\bt|-\rgl$ for
qubit $a$, and the latter are, in turn, obtained by applying the unitary
operators $H$ and $HZ$, respectively, to the state $\al |0\rgl + \bt
|1\rgl$. With this as a hint one can work out the projectors for which
\eqref{eq17} will be satisfied; they are 
\begin{equation}
    P^m_{\lm5} = H[p^m_\lm]H \ot [+]\ot I + HZ[p^m_\lm]ZH \ot [-]\ot I
\label{eq25}
\end{equation}
for $m=0$ and 1, and are obviously located in the pair $a,b$.
As this is the case for any initial basis $\lm$, it shows that the channel
itself is located in (or on) the pair of qubits $a$ and $b$.  

	A similar analysis shows that the same is true at time $t_4$, and this
is not surprising, since between $t_4$ and $t_5$ the only thing that happens is
a unitary transformation on $a$, and while this can change the form in which
the information is located on the two qubits, it cannot affect its presence.
Indeed,  the same sort of reasoning shows that one should not be surprised that
the information is in $a,b$ at $t_4$, since it is already in $a$ and therefore
$a,b$ at $t_3$, and between $t_3$ and $t_4$ these qubits interact with each
other, but with nothing else.  But is it really true that the
information is in the $a,b$ pair at $t_3$ given that it is in $a$?  Yes,
because the projectors used to establish the latter are given in 
\eqref{eq24}, and the projector $[p^m_\lm]\ot I\ot I$ on
$a$, is also a projector $([p^m_\lm]\ot I)\ot I$ on $a,b$.  Note that the
argument by which the information initially in $a$ at $t_3$ should be present
in the pair $a,b$ at $t_4$ depends upon the fact that the quantum gates in the
circuit represent unitary, and thus reversible, operations; one would not reach
the same conclusion if they represented stochastic processes.  Indeed if one
adopts, as is quite possible, frameworks in which some gates produce random
effects, information can disappear.  (See the first example of Sec.~\ref{sct7c},
where the action of the Hadamard gate acting on qubit $a$ destroys
information.)

	Another way of seeing that the channel must be in $a,b$ at $t_5$ is to
note that it can be extracted and placed once again in qubit $a$ by replacing
the part of the circuit following $t_6$ in Fig.~\ref{fg2} with a different set
of gates acting only on $a,b$.  One choice is simply to invert the unitary
operation produced earlier on $a,b$ between $t_3$ and $t_5$ by applying a
second Hadamard gate to $a$, followed by another controlled-not between $a$
(control) and $b$.

	It is immediately evident from the first expression for $|\Psi_5\rgl$
in \eqref{eq22}, and also from the second expression if one multiplies out the
products, that this state is unchanged if one interchanges $b$ and $c$.
Consequently, the argument based on \eqref{eq17} for the presence of the
channel in $a,b$ can also be used to show that it is present in $a,c$: one only
needs to interchange the second and third projectors in each of the terms on
the right side of \eqref{eq25}.  Although the presence of the channel in $a,c$
is not as intuitively obvious as in the case of $a,b$, one can still check that
it is correct by once again replacing that part of the circuit in
Fig.~\ref{fg2} which follows $t_6$ with a suitable set of gates involving only
qubits $a$ and $c$ in such a way that the channel re-emerges in $a$ (or, if one
prefers, in $c$); this construction is left as an exercise.

	But how can it be that the channel is present in $a,b$ and also present
in $a,c$, but not present in $a$ alone?  Here one needs to remember that
information is an abstract entity, not a physical object, and the rules for its
location  do not have to satisfy the axioms of set theory.  If
there is a mystery here, it is not a quantum mystery, for the same thing occurs
in the following classical example.  Let 0 be encoded as $RRR$ or $GGG$, with
equal probability, on three colored slips of paper $abc$, and let 1 be encoded
as $RGG$ or $GRR$, again with equal probability.  The information which
distinguishes 0 from 1 is not present in any of the slips of paper separately,
as in both cases $R$ and $G$ occur with equal probability.  However, it can be
obtained by comparing $a$ with $b$ or $a$ with $c$, whereas comparing $b$ with
$c$ tells one nothing.

	There is, nonetheless, at $t_5$ a nonclassical effect in that for most
choices of the initial basis $\lm$ (the exceptions are $\lm=0$ and $|\lm|=1$),
the framework needed to exhibit the presence of information in the pair $a,b$
is incompatible with that needed for the pair $a,c$.  Consequently, while it
makes sense to say that the \emph{channel} is in $a,b$, or that it is in $b,c$,
it does not make sense to say that it is both in $a,b$ \emph{and} in $a,c$.

	\subsection{Alternative frameworks}
\label{sct7c}

	Just as in the case of dense coding, Fig.~\ref{fg1}, additional
insight into the circuit in Fig.~\ref{fg2} can be obtained by considering what
happens in some special frameworks. In the first of these the $S_z$ or
computational basis is employed for all three qubits at all times.  This
resembles framework $\FS_2$ in the case of dense coding, Sec.~\ref{sct5b}, so
it will suffice to note only the main points.  At $t_2$ (and therefore $t_3$)
the $b$ and $c$ qubits are either both 0 or both 1, with equal probability, and
thus at $t_4$ the information about whether $a$ was 0 or 1 at $t_0$ resides in
correlations between $b$ and $c$ --- they are equal for $a=0$, and different if
$a=1$ --- as well as in qubit $a$ itself.  But at $t_5$ the Hadamard gate has
randomized $a$, and it is uncorrelated with its earlier value, whereas that
information is still present in the pair $b,c$.  In particular, at $t_5$ the
information is not present in the pair $a,b$, nor is it present in $a,c$.  This
does not contradict the result of Sec.~\ref{sct7b}, for the presence or absence
of information in some particular location depends upon the framework used to
describe the quantum system, and the one under discussion is incompatible with
the frameworks needed to exhibit the same information in $a,b$ or in $a,c$.
The presence of information in correlations between $b$ and $c$ is limited to
the $\lm=0$ or $S_z$ basis for qubit $a$ at $t_0$; for all other $\lm$ the
information is, at best, partially present.  Next, the controlled-not between
$t_6$ and $t_7$ writes the information from the $b,c$ correlation into qubit
$c$, while in this framework the final controlled-Z gate has no effect, as in
the analogous situation in Fig.~\ref{fg1}.

	The second framework we will consider employs the $\lm=1$ or $S_x$
basis for $a$ at $t_0$, and also for this and the other qubits at all later
times, except that the $S_z$ (computational) basis is used for $a$ at times
$t\geq t_5$. The situation resembles framework $\FS_3$ in the case of dense
coding, Sec.~\ref{sct5b}, in that at $t_2$ (and $t_3$) the $b,c$ pair is with
probability 1/2 in the state $++$ and with probability 1/2 in $--$. The action
of the controlled-not gate acting just after $t_3$ is to leave the $a$ qubit
($+$ or $-$) unchanged if $b=+$, or flip it from $+$ to $-$ or vice versa if
$b=-$; in either case $b$ remains unchanged. Consequently, at $t_4$ the
information $a=+$ or $-$ originally present at $t_0$ is no longer present in
$a$ by itself, but in a correlation between $a$ and $b$, or, equivalently,
between $a$ and $c$.  The situation is very similar to the classical analogy
discussed above in Sec.~\ref{sct7b}.  The Hadamard gate changes $a$ from $+$ to
0 or $-$ to 1 between $t_4$ and $t_5$, so if we use the $S_z$ basis for $a$ at
$t_5$ and later times the initial information about $a$ at $t_0$ continues to
be present in the correlations between $a$ and $c$.  The next controlled-not
gate in Fig.~\ref{fg2} changes the $b$ qubit to $+$, so it no longer contains
any information, and the final controlled-Z transfers the information in the
correlation between $a$ and $c$ to $c$, where it is located at $t_8$.

	\subsection{Teleportation}
\label{sct7d}

	As noted in Sec.~\ref{sct7a}, in the usual teleportation scenario Alice
measures qubits $a$ and $b$ in Fig.~\ref{fg2} in the $S_z$ or computational
basis at $t_5$ and communicates the results to Bob over a classical channel.
The same same result can be obtained, following the proposal in
\cite{BrBC98} but using somewhat different language, by employing a
structure of frameworks in which qubits $a$ and $b$ are in one of the $S_z$
eigenstates for $t\geq t_5$, but (in contrast to the first example in
Sec.~\ref{sct7c}) $c$ is not.  Instead, at $t_5$ one uses an orthonormal basis
\begin{equation}
    |00 p^m_\lm\rgl,\quad |01 q^m_\lm\rgl,\quad |10 r^m_\lm\rgl,\quad
  |00 s^m_\lm\rgl
\label{eq26}
\end{equation}
for the system of three qubits, where $m$ is 0 or 1, $|p^m_\lm\rgl$ is defined
in \eqref{eq14}, and the states $|q^m_\lm\rgl$, $|r^m_\lm\rgl$, $|s^m_\lm\rgl$
are obtained from $|p^m_\lm\rgl$ by applying the unitary transformations $X$,
$Z$, and $XZ$, respectively.  This basis consists of product states, but the
$c$ states are ``contextual'' or ``dependent'' in the notation of \cite{Grf02},
Ch.~14.  (The term ``nonlocal'' employed in \cite{Beal99} is, in our opinion,
somewhat misleading.)

	If we use the $\lm$ basis for $a$ at $t_0$, the basis states in
\eqref{eq26} at $t_5$, and the images of the states in \eqref{eq26} under
unitary time development at later times, the result is a framework in which the
initial information, $m=0$ vs.\ $m=1$ in the $\lm$ basis at $t_0$, is contained
in correlations between all three qubits at time $t_5$ and $t_6$, in
correlations between $a$ and $c$ at $t_7$, and in qubit $c$ at $t_8$.  That all
three qubits are needed at $t_5$ does not contradict the fact,
Sec.~\ref{sct7b}, that for each $\lm$ there is an alternative framework in
which the information is contained in qubits $a$ and $b$; it is merely one more
example of the dependence of the location of information upon the choice of
framework.  In the same way, the location of a quantum channel depends upon the
structure employed for describing it: If one uses the structure under
discussion, the collection of frameworks that employ
\eqref{eq26} at $t_5$, then at $t_5$ the channel is located in or expressed as
correlations among all three qubits, and not in any smaller subset. Indeed, if
one uses the $S_z$ basis for $a$ and $b$ at $t_5$ there is \emph{no}
information whatsoever in these two qubits about the initial state for
\emph{any} choice of $\lm$; their values are completely random, in accord with
the usual statement that Alice's measurements tell her nothing about the nature
of the state that is being teleported.

	Note that by assuming that the two qubits $a$ and $b$ are sent from
Alice to Bob through a quantum channel, as suggested by Fig.~\ref{fg2}, but at
the same time requiring that they be described in the $S_z$ basis for all
frameworks in the structure, one arrives at a situation whose end result is
just the same as if Alice had measured $a$ and $b$ in the $S_z$ basis and the
results had been sent to Bob over a classical channel in order to actuate the
final two gates acting on $c$.  The circuit in Fig.~\ref{fg2} is not an
improvement over the original teleportation scheme in a commercial sense, for
one of the motivations for the latter was to remove any need for a quantum
channel between Alice and Bob once the initial entangled state has been
established, i.e., at any time after $t_3$.  However, both ``measurement'' and
``classical communication'' are somewhat complicated ideas, and replacing them
with alternative notions which apply to a simple quantum circuit without the
need to invoke external apparatus might lead to conceptual simplifications in
other applications of quantum information theory.

	\section{Conclusion}
\label{sct8}

	This paper proposes a way in which to view, or analyze, or think about
quantum information, and, as indicated towards the end of Sec.~\ref{sct1}, it
is convenient to summarize it in terms of two theses, one of which seems well
founded, though it may in the end turn out to be defective, while the other is
more speculative. The first and less controversial thesis is that a framework
or consistent family, Sec.~\ref{sct4}, provides a natural means of extending
classical information theory, both its formalism and the associated intuition,
into the quantum domain.  This is because a framework provides a consistent
probabilistic description of a quantum system in which probabilities have all
their usual properties (they are positive, sum to 1, etc.) and their usual
intuitive interpretation. The only difference is that now they apply to quantum
properties described by a quantum Hilbert space.  There is no other consistent
way of embedding standard probability theory in standard quantum mechanics
(using a Hilbert space without hidden variables) known at the present time,
aside from the well-known approach based on measurements, whose inadequacy for
the purposes of this paper was pointed out in Sec.~\ref{sct4a}.  In any case,
all the probabilities for measurement outcomes which one can obtain through a
valid application of the (not always very clear) rules in the textbooks are
also consequences of the correct application of histories methods using
frameworks, so the former approach is subsumed under the latter.

	The example of dense coding, Sec.~\ref{sct5}, shows that classical
ideas about information of the sort discussed in Sec.~\ref{sct2} can be
translated in a fairly natural way into the quantum domain, and the results are
physically sensible as long as one adheres to the single framework rule,
something which is in any case necessary for a consistent interpretation of
quantum mechanics; see the discussion in Sec.~4.6
of \cite{Grf02}.  That dense coding is able to transport two bits of
information has nothing to do with mysterious long-range influences, which are
in any case absent from quantum theory when properly formulated \cite{nt9}, 
or with information traveling backwards in time, and it is fully consistent
with the condition on the maximum capacity of a quantum carrier of information
(no more than one bit per qubit) stated in Sec.~\ref{sct4b}.  Its peculiarity
arises from the fact that the tensor product of two quantum Hilbert spaces
provides a very different description of a physical system than does the
Kronecker product of two classical phase spaces.  The analysis of the
teleportation circuit in Sec.~\ref{sct7} using a variety of frameworks provides
additional examples of how this approach can answer questions about the
physical location of information and its flow from one region in space to
another.  In addition, it is possible to argue that many of the ideas in the
published literature having to do with ``classical'' information in the quantum
context, such as the classical capacity of a quantum channel, have a quite
natural formulation in terms of frameworks.

	Further support for the first thesis comes from the following
observation.  Quantum physicists (with some notable exceptions) believe that
the macroscopic, classical world is also governed by quantum principles at a
fundamental level, so that classical mechanics is a particular limiting case of
quantum mechanics; see \cite{Wil00} for one expression of this widespread
faith.  The histories approach provides a program, not yet complete but very
plausible \cite{nt11}, for understanding classical physics in precisely this
way, by the use of quasiclassical frameworks.  The first thesis states, in
essence, that quantum information theory consists in applying the ideas of
classical information theory to quantum processes described by a single
(quasiclassical) framework.  We can then ask what the implications of quantum
information theory, as defined in this way, are for the classical or
macroscopic world.  The answer is quite clear: because the classical world can
be described quantum mechanically using a single framework, quantum information
theory in the classical context is automatically the same as standard classical
information theory.  While such a ``correspondence principle'' is obviously not
a proof that quantum information can be correctly formulated in the way we are proposing, it
does provide some support.

	We now come to the second thesis, which is much more speculative. It is
that \emph{all} of quantum information theory is, ultimately, just a matter of
applying ``classical'' information theory in different frameworks to a quantum
system, and paying attention to framework dependence.  That is, there is no
special form of ``quantum'' information lying outside the purview encompassed
by the first thesis.  To put it in another way, the second thesis claims that
the concept of information in a particular framework, together with a
consideration of various collections of frameworks (thus structures, in the
sense defined in Sec.~\ref{sct6a}) is an adequate tool for formulating the
various problems, such as entanglement, cryptography, and the capacity of
quantum channels, which nowadays constitute the central concerns of quantum
information theory.  Being able to formulate the problems in this way does not
mean that they will be easy to solve, for, as anyone working in the field is
well aware, there are a host of formidable technical difficulties confronting
attempts to do calculations even in systems as simple as  two qubits.
Nonetheless, being able to formulate the different problems from this
perspective, if it is possible, could provide a certain unity or coherence to
quantum information theory, something which has been lacking up to now. 

	The first ``test case'' for the second thesis is the analysis of the
teleportation circuit in Sec.~\ref{sct7}, carried out using various different
frameworks.  In essence the circuit represents a particular form of quantum
channel, thus capable of transmitting quantum information, whatever that may
be.  Is there anything essential missing from our discussion based upon
structures consisting of collections of frameworks, each using a different
basis for the channel input?  Is there any physical question which cannot be
addressed in this manner?  It is clear that there are various interesting
questions that have not yet been addressed from this point of view, such as why
it is that one needs two ``classical'' bits in the standard teleportation
process, and it may turn out that the frameworks approach lacks adequate
concepts to deal with this and similar issues.  Only further research will show
whether the use of frameworks can provide a reasonably complete understanding
of teleportation and of
other significant problems of quantum information theory.  While the analysis
of the teleportation circuit represents an encouraging first step, much remains
to be done.

	Thus I present the second thesis not as something for which there are
strong and compelling arguments, but rather as an idea worth exploring.  Even
if partly successful it could prove to be a significant advance in our
understanding of quantum systems in information-theoretic terms, and if it
fails, this itself could be the source of interesting insights.

	\section*{Acknowledgments}

	The research described here was supported by the
National Science Foundation Grant PHY 99-00755.

%	\appendix* \section{Deutsch and Hayden} %revtex
	\section*{Appendix: Deutsch and Hayden} %latex
\label{apa}

	Deutsch and Hayden in \cite{DtHy00} use the following approach to
describe quantum information in a collection of $n$ qubits undergoing unitary
time evolution, where we use a slightly different notation than theirs.  Let
$\sg^\al_j$ be the $j$'th Pauli spin operator for qubit $\al$, where $j=x,y$,
or $z$, and $\al = 1,2,\ldots n$, and let
\begin{equation}
\hat \sg^\al_j(t) = T(0,t) \sg^\al_j T(t,0),
\tag{A.1} %comment out for revtex
\label{eqa1}
\end{equation}
with $T$ the time development operator in \eqref{eq10}, be the corresponding
Heisenberg operator at a reference time $t=0$, as a function of the (physical)
time $t> 0$.  Deutsch and Hayden adopt the view that the three operators
$\{\hat \sg^\al_j\}$ for $j=x,y,z$ constitute a quantum description of qubit
$\al$ at time $t$, and the collection of all such operators, for $\al =1,
2,\ldots n$ provide a complete description of the entire system at time $t$.
Note that the $\{\hat \sg^\al_j\}$ are (in general rather complicated)
operators on the full $2^n$-dimensional Hilbert space.  Probabilities are
expressed, as is usual in the Heisenberg picture, by using expectation values
of operators of the form \eqref{eqa1}, products of these for different $\al$,
and sums of such products, in a suitable initial state, which is chosen to be
$|0\rgl\ot|0\rgl\cdots |0\rgl$.

	The unitary time transformation $T$ may depend on a real parameter
$\th$.  For example, during the interval from $t_1$ to $t_1+\Dl t$ qubit $5$
may pass through a unitary gate corresponding to a rotation of the Bloch sphere
by an angle $\th$ around some axis. As a consequence, at later times some of
the $\hat \sg^\al_j$, in particular the $\hat \sg^5_j$, but also Heisenberg
operators for other qubits if they have interacted with this one after the time
$t_1+\Dl t$, will depend upon $\th$.  \emph{Information about} $\th$ is, by
definition, said to be located \emph{in} the qubits whose Heisenberg operators
depend upon $\th$.  Even if at time $t$ one or more components of $\hat
\sg^\al_j(t)$ for a given $\al$ depend upon $\th$, it is possible that the
probabilities associated with any measurement carried out on this qubit at time
$t$ will be independent of $\th$; see \cite{DtHy00} for examples. In such a
case, while the information is present in this qubit, it is
\emph{inaccessible}.  The same ideas apply to cases in which $T$ depends upon
additional parameters $\phi$, $\chi$, etc.  Information defined in this way has
a definite location and flows from place to place through interactions among
the qubits: there is never any instantaneous or superluminal transfer between
non-interacting subsystems.
	
	When applied to the examples considered in the present paper, the
Deutsch and Hayden prescription would say that in the case of dense coding, the
two bits of information sent by Alice to Bob are present but inaccessible in
the qubit which see sends him, during the interval between $t_5$ and $t_6$ in
Fig.~\ref{fg1}, rather than, as we argued in Sec.~\ref{sct5}, in the
correlations between this qubit and the one already in Bob's possession. In the
case of teleportation the information initially present in the unknown state
$|\psi\rgl$ of qubit $a$ in Fig.~\ref{fg2} is in the (classical) two-bit signal
sent from Alice to Bob after her measurement, rather than in the correlation
between these and the qubit in Bob's possession, as we argued in
Sec.~\ref{sct7}.  There are various other differences between the conclusions
reached by Deutsch and Hayden and those in the present paper. One of particular
interest, as it involves a purely ``classical'' situation, is the third example
in Sec.~\ref{sct2}, where they would argue that the information which Alice
sends to Bob is present (but, once again, inaccessible) in the single slip of
paper she sends to him, whereas we argued that it is not present there,
but instead in a statistical correlation between this and a different slip
of paper. 

	The differences between the Deutsch and Hayden approach and that found
in the present paper reflect two quite distinct approaches to defining
``quantum information.''  The first is based on a notion of information as
reflecting causal influences, whereas the second uses statistical correlations.
The latter seems closer to the perspective of classical information theory as
developed by Shannon and his successors, but of course this does not imply that
it is the correct, much less the unique approach to use when dealing with
quantum systems.  The one point at which the two approaches agree is in
affirming that, just as in the case of classical information, quantum
information cannot be transmitted instantly between non-interacting systems;
for more on this from the perspective of the present paper, see Chs.~23 and 24
of \cite{Grf02}.  However, this agreement is only superficial, since Deutsch
and Hayden assert that a quantum description of the state of each individual
qubit is possible even when the system as a whole is in an entangled state, in
contrast to the discussion of Bell states found in Sec.~\ref{sct4c} above.  The
reader is invited to compare these approaches and make up his own mind about
their virtues and vices, preferably after a careful reading of \cite{DtHy00},
since the summary given above is necessarily very brief, and the original paper
contains several helpful examples along with much more detail.


\begin{thebibliography}{10}
\bibitem{nt2}
See p.~163 of \cite{Omn94}, where a ``consistent logic'' is what we call a
  ``framework'', Sec.~III of \cite{Grf96}, Sec.~16.1 of \cite{Grf02}.
\bibitem{nt7}
This equivalence is most easily understood by noting that ideal measurements
  always reveal a pre-existing state of affairs if one adopts an appropriate
  framework, see the discussion in Ch.~17 of \cite{Grf02}, and by employing for
  the circuit in Fig.~\ref{fg2} a framework in which the pair $a,b$ is in one
  of the Bell states at $t_3$ and in one of the computational basis states at
  $t_5$.
\end{thebibliography}
\end{document}